\documentclass[fleqn,usenatbib]{mnras}

\usepackage[T1]{fontenc}
\usepackage{ae,aecompl}

\usepackage{graphicx}	
\usepackage{amsmath}	
\usepackage{amssymb}	
\usepackage{multicol,multirow}
\usepackage{booktabs,tabularx}

\newcommand{\ee}[1]{\times 10^{#1} } 
\newcommand{\hst}{\textit{HST}} 
\newcommand{\spitzer}{\textit{Spitzer}} 
\newcommand{\msun}{\text{M}_\odot} 
\newcommand{\lsun}{\text{L}_\odot}

\title[LBT search for failed supernovae]{The search for failed supernovae with the Large Binocular Telescope: a new candidate and the failed SN fraction with 11 yr of data}

\author[Neustadt et al.]{
J.~M.~M.~Neustadt,$^{1}$\thanks{E-mail: neustadt.7@osu.edu (JMMN)}
C.~S.~Kochanek,$^{1,2}$ 
K.~Z.~Stanek,$^{1}$
C.~Basinger,$^{1}$
\newauthor{
T.~Jayasinghe,$^{1}$
C.~T~.~Garling,$^{1}$
S.~M.~Adams,$^{1}$
J.~Gerke$^{1}$
}
\\
$^{1}$Department of Astronomy, The Ohio State University, 140 West 18th Avenue, Columbus, OH 43210, USA \\
$^{2}$Center for Cosmology and AstroParticle Physics (CCAPP), The Ohio State University, 191 W. Woodruff Avenue, Columbus, OH 43210, USA \\ 
}

\date{Accepted XXX. Received YYY; in original form ZZZ}

\pubyear{2019}

\begin{document}
\label{firstpage}
\pagerange{\pageref{firstpage}--\pageref{lastpage}}
\maketitle

\begin{abstract}
We present updated results of the Large Binocular Telescope Search for Failed Supernovae.  This search monitors luminous stars in 27 nearby galaxies with a current baseline of 11~yr of data. We re-discover the failed supernova (SN) candidate N6946-BH1 as well as a new candidate, M101-OC1. M101-OC1 is a blue supergiant that rapidly disappears in optical wavelengths with no evidence for significant obscuration by warm dust.  While we consider other options, a good explanation for the fading of M101-OC1 is a failed SN, but follow-up observations are needed to confirm this.  Assuming only one clearly detected failed SN, we find a failed SN fraction $f = 0.16^{+0.23}_{-0.12}$ at 90 per~cent confidence.  We also report on a collection of stars that show slow ($\sim$decade), large amplitude ($\Delta L/L > 3$) luminosity changes.
\end{abstract}


\section{Introduction}\label{intro}

The life of every massive star (>8~$\msun$) ends with the collapse of its core, which is sometimes followed by the violent ejection of its envelope and the production of a luminous core-collapse supernova (ccSN).  Volume-limited samples have shown that around 59 per~cent of ccSNe are the hydrogen-rich Type II-P class, with their progenitors identified as red supergiants (RSGs) \citep{smartt09,li11}.  However, the most massive of the observed Type II-P progenitors has a mass of only 16--18~$\msun$ \citep{smartt09,smartt15}, which is significantly lower than the maximum red supergiant mass of $\sim$25~$\msun$ \citep{humphreys79}.  This implies that the more massive RSGs are not exploding as SNe.  The alternative hypothesis, which supposes that increased RSG mass loss causes the stars to explode as stripped SNe (e.g., \citealt{ekstrom12,groh13}), requires higher mass loss rates than observed (e.g., \citealt{beasor20}).

Modern theoretical models find that these ``missing'' RSGs with masses of 18--25~$\msun$ are forming cores that are too compact to explode as SNe \citep{oconnor11,ugliano12,sukhbold14,pejcha15,ertl16,sukhbold16}.  These observational and theoretical findings have led to the hypothesis that these massive RSGs instead become ``failed SNe'' \citep{kochanek08}, and form black holes (BHs) with masses of $\sim$5--10$~\msun$ that are typical of those observed in our Galaxy \citep{kochanek14,kochanek15}.  

The origin of BHs is particularly important in the wake of the BH-BH and BH-neutron star mergers being detected by the Laser Interferometer Gravitational Wave Observatory (LIGO, \citealt{abbott16-ligo}).  These mergers are the end result of a long and complicated process involving binary stellar evolution and the core-collapse of each progenitor star (e.g., \citealt{abbott16,belczynski16,woosley16}).  Understanding the formation of BHs and their subsequent evolution is crucial to understanding the mergers detected by LIGO and future gravitational wave observatories.  Here we will focus on detecting the formation of BHs, but other relevant probes include searches for non-interacting, low-mass BHs \citep{thompson19,jayasinghe21}, searches and constraints on binaries, bound and unbound, in SN remnants (e.g., \citealt{kochanek18,kerzendorf19,kochanek19,maitra19}), and searches for BHs using gravitational microlensing (e.g., \citealt{lu16}). 

Here we present updates on the search for failed SNe with the Large Binocular Telescope (LBT), first proposed by \citet{kochanek08}.  This survey monitors luminous stars in 27 galaxies within 10~Mpc using the Large Binocular Telescope (LBT) and is designed to detect the death of evolved $\sim$9--30~$\msun$~stars, independent of whether they explode as SNe.  We reference the previous papers in the survey, \citet{gerke15}, \citet{adams17a}, \citet{adams17b}, and \citet{basinger20}, as G15, A17a, A17b, and B20, respectively.  G15 put the first constraints on the failed SN fraction as derived from the first 4~yr of data and identified a first candidate failed SN, A17b updated the constraints using 7~yr of data, and A17a led a detailed analysis of the failed SN candidate N6946-BH1.  The most recent analysis of the late-time evolution of N6946-BH1 can be found in B20.

N6946-BH1 is the best existing candidate for a failed SN. It was a $\sim$10$^{5.5}~\lsun$, $\sim$25$~\msun$ RSG that underwent a luminous $\sim$10$^{6}~
\lsun$ transient before mostly vanishing in optical and mid-IR wavelengths (A17a).  The transient was likely powered by the recombination of the RSG's envelope that would be gently ejected after core-collapse \citep{nadezhin80,lovegrove13}. Follow-up photometry with the \textit{Hubble Space Telescope} (\hst) and the \textit{Spitzer Space Telescope} (\spitzer) showed the existence of a faint $\sim$2000$~\lsun$ remnant luminosity that could be understood as emission from fallback accretion (A17a, B20).

Other surveys have also searched for failed SNe.  \citet{reynolds15} examined a sample of 15 galaxies with multi-epoch \hst~data and reported the discovery of NGC~3021-CANDIDATE-1, a 25--30~$\msun$ yellow supergiant (YSG), which disappeared in the optical without a recorded SN.  Another failed SN candidate is a luminous blue variable (LBV) in the dwarf galaxy PHL~293B \citep{allan20}.  While N6946-BH1 and NGC~3021-CANDIDATE-1 were identified by their photometry, this LBV was identified as a failed SN candidate due to the disappearance of broad emission lines, which had been present in the spectra of the object for years prior.  While one interpretation is that this source is a failed SN, it is also posited that the source is instead the remnant of an undetected Type IIn SN, or perhaps some other type of stellar variability \citep{burke20}. 

In addition to constraints on the failed SN fraction, the LBT survey has also produced results on SNe progenitors \citep{johnson17}, SN progenitor variability \citep{szczygiel12,kochanek17,johnson18}, ``SN imposters'' \citep{adams15}, SN~2008S and similar events \citep{adams16,andrews20}, and LBVs \citep{grammer15}. In particular, \citet{johnson18} showed that the typical Type~II progenitor has no pre-SN mass ejection phase.

In this paper, we present updated results on the failed SN fraction following the work done in G15 and A17b using 11~yr of data.  This represents an increase in the baseline of the survey by up to 4~yr compared to the previous results from A17b.  In Sections \ref{sec:im_sub} and \ref{sec:can_sel}, we discuss our image subtraction methods and our criteria for selecting failed SN candidates from among the millions of processed light curves. In Section \ref{sec:sne}, we discuss our observations of SNe and other luminous transients that occur in our survey.  In Section \ref{sec:candidates}, we discuss two new ``disappearing stars'', one of which we classify as a failed SN candidate and one which we ultimately reject. In Section \ref{sec:weird}, we present a sample of peculiar large-amplitude, slowly-varying stars.  In Section \ref{sec:conclusion}, we summarize our work and update our estimates of the failed SN fraction.

\section{Image Subtraction}\label{sec:im_sub}

\begin{table*}
\caption{Sample of galaxies in the LBT survey}\label{tab:sample}
\begin{tabularx}{\textwidth}{l *{1}{>{\centering\arraybackslash}X} c c *{3}{>{\centering\arraybackslash}X}} \toprule
\multicolumn{1}{c}{Galaxy} & Number of & \multicolumn{2}{l}{\hphantom{00}Observation period} & Baseline & Distance & Distance \\
 & epochs & Start & End & (yr) & (Mpc) & reference \\ \midrule
M81/NGC~3031 & 45 & 2008-03-08 & 2019-10-26 & 11.1 & 3.65 & 1 \\
M82/NGC~3034 & 34 & 2008-03-08 & 2019-10-26 & 11.1 & 3.52 & 2 \\
M101/NGC~5457 & 35 & 2008-03-08 & 2019-12-21 & 11.1 & 6.43 & 3 \\
NGC~628/M74 & 27 & 2008-11-22 & 2019-12-20 & \leavevmode\hphantom{0}9.8 & 8.59 & 4 \\
NGC~672 & 26 & 2008-07-05 & 2019-12-22 & 10.0 & 7.2\leavevmode\hphantom{0} & 5 \\
NGC~925 & 26 & 2008-07-06 & 2019-12-21 & 10.0 & 9.16 & 6 \\
NGC~2403 & 44 & 2008-05-05 & 2019-10-26 & 10.2 & 3.56 & 7 \\
NGC~2903 & 19 & 2008-03-08 & 2019-12-21 & 11.5 & 8.9\leavevmode\hphantom{0} & 8 \\
NGC~3077 & 25 & 2008-05-04 & 2019-10-26 & 10.4 & 3.82 & 5 \\
NGC~3344 & 21 & 2008-05-04 & 2019-03-31 & \leavevmode\hphantom{0}9.5 & 6.9\leavevmode\hphantom{0} & 9 \\
NGC~3489 & 20 & 2008-03-12 & 2019-12-21 & 10.9 & 7.18 & 10\leavevmode\hphantom{0} \\
NGC~3623/M65 & 22 & 2008-05-04 & 2019-03-31 & \leavevmode\hphantom{0}9.3 & 10.62\leavevmode\hphantom{0} & 11\leavevmode\hphantom{0} \\
NGC~3627/M66 & 22 & 2008-05-04 & 2019-03-31 & \leavevmode\hphantom{0}9.3 & 10.62\leavevmode\hphantom{0} & 11\leavevmode\hphantom{0} \\
NGC~4214 & 19 & 2008-03-13 & 2019-04-01 & \leavevmode\hphantom{0}9.2 & 2.98 & 12\leavevmode\hphantom{0} \\
NGC~4236 & 18 & 2008-03-09 & 2019-03-30 & \leavevmode\hphantom{0}9.3 & 3.65 & 1 \\
NGC~4248 & 44 & 2008-03-08 & 2019-03-31 & 10.2 & 7.21 & 13\leavevmode\hphantom{0} \\
NGC~4258/M106 & 44 & 2008-03-08 & 2019-03-31 & 10.2 & 7.21 & 13\leavevmode\hphantom{0} \\
NGC~4395 & 17 & 2008-03-10 & 2019-03-31 & \leavevmode\hphantom{0}7.1 & 4.27 & 14\leavevmode\hphantom{0} \\
NGC~4449 & 22 & 2008-03-09 & 2019-12-21 & 10.2 & 3.82 & 15\hphantom{0} \\
NGC~4605 & 18 & 2008-03-13 & 2019-03-31 & \leavevmode\hphantom{0}9.1 & 5.47 & 16\leavevmode\hphantom{0} \\
NGC~4736/M94 & 17 & 2008-03-10 & 2019-03-31 & \leavevmode\hphantom{0}9.3 & 5.08 & 17\leavevmode\hphantom{0} \\
NGC~4826/M64 & 21 & 2008-03-08 & 2019-03-30 & 10.0 & 4.4\leavevmode\hphantom{0} & 2 \\
NGC~5194/M51 & 25 & 2008-03-09 & 2019-03-31 & \leavevmode\hphantom{0}9.4 & 8.3\leavevmode\hphantom{0} & 18\leavevmode\hphantom{0} \\
NGC~5474 & 21 & 2008-03-13 & 2019-03-31 & \leavevmode\hphantom{0}9.4 & 6.43 & 3 \\
NGC~6503 & 29 & 2008-05-04 & 2019-10-24 & 11.2 & 5.27 & 6 \\
NGC~6946 & 48 & 2008-05-03 & 2019-10-24 & 10.4 & 5.96 & 19\leavevmode\hphantom{0} \\
IC~2574 & 24 & 2008-03-13 & 2019-12-21 & 10.9 & 4.02 & 6 \\ \bottomrule
\end{tabularx}
\begin{flushleft} \textit{Notes}: The baseline is the time from the second observation to the penultimate observation in the selection period. References: (1) \citet{gerke11}; (2) \citet{jacobs09}; (3) \citet{shappee11}; (4) \citet{hermann08}; (5) \citet{karachentsev04}; (6) \citet{karachentsev03}; (7) \citet{willick97}; (8) \citet{drozdovsky00}; (9) \citet{verdes00}; (10) \citet{theureau07}; (11) \citet{kanbur03}; (12) \citet{dopita10}; (13) \cite{herrstein99}; (14) \citet{thim04};(15) \citet{annibali08}; (16) \citet{karachentsev06}; (17) \citet{tonry01}; (18) \citet{poznanski09}, and (19) \citet{karachentsev00}. \end{flushleft}
\end{table*}

We follow the methods described in G15 and A17b. We use the \texttt{ISIS} image subtraction package \citep{alard98,alard00} with the same astrometric references as those used in G15.  We updated the reference images used for image subtraction by including higher-quality data collected over the current $\sim$11~yr baseline of the survey.  These references images had better full-widths at half maximum (FWHMs) as well as higher S/N than those used by A17b.  These updates were especially important for the \textit{UBV} images where the number of epochs used to construct the reference images doubled for some fields.  We ran image subtraction on all epochs to construct light curves, but for light curve analysis/candidate selection, we excluded epochs with FWHM~>~2~arcsec to exclude epochs with bad seeing, background counts >~30,000 to exclude observations taken during twilight, and image subtraction scaling factors <~0.4 to exclude observations with significant cirrus.  The number of included/analyzed epochs is given in Table \ref{tab:sample}.  For candidate selection, we only used data taken before January 2020, but later data were used to evaluate candidates.

We follow the same methods for calibration as used by G15 and A17b.  Sloan Digital Sky Survey (SDSS, \citealt{ahn12}) stars with SDSS \textit{ugriz} AB magnitudes are matched with stars in the reference images and transformed to \textit{UBVR} Vega magnitudes using the conversions reported by \citet{jordi06} and zero-points reported by \citet{blanton07}.  For the fields where SDSS stars were unavailable, we followed the prescriptions described in G15. The $U$-band data for IC~2574, NGC~925, and NGC~6946 remain uncalibrated. 

As in A17b, we use mask files for saturated pixels as opposed to actually masking the saturated pixels in the images by changing the pixel value.  This allows us to keep track of sources that were saturated in some epochs but not all.  We use the subtracted images generated by \texttt{ISIS} to construct a root mean square (RMS) image.  This combines the subtracted images such that each pixel in the RMS image is the RMS of that position's pixel values in all the subtracted images.  This process highlights variable sources.

\section{Candidate Selection}\label{sec:can_sel}

For each field, we generate a master catalog of sources by combining two catalogs: (1)  a catalog of `bright sources' ($\nu L_\nu > 1000~\lsun$) generated by running \texttt{DAOPhot} \citep{stetson87} on the reference image; and (2) a catalog of `RMS sources' generated by running \texttt{SEXTRACTOR} \citep{bertin96} on the RMS image.  Throughout the paper, $\nu L_\nu$ refers to the band luminosity associated with a given filter while $L$ refers to the bolometric luminosity.  For a magnitude corresponding to a flux $F_v$, $\nu L_\nu = 4\pi D_L^2 \nu F_\nu$ where $D_L$ is the luminosity distance.  We will then select targets based on the temporal changes $\Delta \nu L_\nu$ in these band luminosities.  As an example of a field, M101 in $R$-band had 235,559 bright sources, 9,001 RMS sources, and 5,185 sources that appeared in both the bright and the RMS catalogs.  Including all galaxies and filters, the master catalog included 6.3~million sources. 
As shown in G15, we should be sensitive to all core-collapse progenitors except for some Wolf-Rayet stars if we search for band luminosity changes $\lvert\Delta\nu L_\nu\rvert >10^4 L_\odot$ in all four filters. The redder filters are better for RSGs and the bluer filters are better for hot, stripped stars. 

From the master catalog, we generate a candidate list by finding sources with light curves that match either of the following two criteria: \\

(1) $\lvert\Delta \nu L_\nu\rvert > 10^4~\lsun$ between all of the following image pairs: first and last, first and penultimate, second and final, and second and penultimate images.  Here, the `second' image is chosen to be at least 1 month after the `first' image.  We also require that the change in flux between the first and last image is greater than 10 per~cent of the flux in the first image.  This criterion is meant to flag sources that become dimmer or brighter over the baseline of the survey, while attempting to exclude variable stars and subtraction artefacts.  

(2) $\nu L_\nu > 10^5~\lsun$ in at least two consecutive epochs and for a period of 3~months to 3~yr.  This criterion is meant to flag transient flares such as those predicted by \citet{lovegrove13} and observed in N6946-BH1 (G15, A17a, B20).  \\

Hereafter we refer to these as criterion 1 and criterion 2.  For all sources that match either criterion, we also require that the source be sufficiently compact in the RMS image.  This is intended to remove spurious RMS sources that are artefacts of image subtraction.  The compactness is computed by comparing the \texttt{SEXTRACTOR} fluxes of the source using apertures of 4 pixels and 8 pixels in radii.  We empirically determined a limit to be $F(4~\text{pix})/F(8~\text{pix}) < 0.3$ for rejecting a source.

In total, we found 13223 sources that satisfied at least one of the selection criteria in any of the filters.  The authors JMMN, CSK, and KZS independently reviewed the light curves and image subtractions of each source.  The vast majority of these sources were bright star image subtraction artefacts (10878, or 82 per~cent) or obvious variable stars (2028, or 15 per~cent), leaving only 317 sources that passed the initial round of inspection.  These 317 sources were matched across filters, leaving 151 distinct sources that were again inspected.  After removing known SNe, peculiar transients, and other residual spurious sources, we finalized our candidate list to five candidates.  Three of these sources are ``rediscovered'' candidates, including N6946-BH1, discussed in earlier papers and in Section~\ref{sec:candidates}, and two are new candidates discussed in Sections~\ref{sec:n4736} and \ref{sec:m101}.

We also re-analyzed the candidate sources presented in A17b.  Most of these were already rejected in A17b as ``slow-faders'' -- objects fading over timescales of >1000~d.  One particular candidate, labelled N925-OC1, was singled out as fading relatively rapidly and was identified as a cool, luminous supergiant.  In our current analysis, we find that in the epochs following those analyzed by A17b, the source returned to near peak brightness in $R$-band, and the source continued to vary over the following $\sim$1200~d.  While it did satisfy criterion 1 in $R$-band for being less luminous in the last two images, the lack of long-term quiescence means we do not consider it a candidate.  

\section{Supernovae and other transients}\label{sec:sne}

\begin{table}
\caption{List of supernovae and other luminous transients}
\label{tab:sne}
\begin{center}
\begin{tabular}{lccc} \toprule
\multicolumn{1}{c}{ID}	& Galaxy & Criteria 	& Classification \\
\midrule
SN~2008S  & NGC~6946 &  1  & IIn \\
SN~2009hd  & NGC~3627 & N/A &  II-L \\
SN~2011dh & NGC~5194  & 1,2 & IIb \\
SN~2012fh  & NGC~3344  & N/A &  Ic \\
SN~2013am & NGC~3623 & 2 & II-P \\
SN~2013ej & NGC~628  & 2 &  II-P \\ 
SN~2014bc & NGC~4258 & 1,2 & II-P \\
SN~2016cok & NGC~3627 & 1,2 & II-P \\
SN~2017eaw & NGC~6946  & 1,2 & II-P \\ \midrule
SN~2011fe & M101 &  2 & Ia \\
SN~2014J & M82  & 1,2 & Ia \\ \midrule
AT2019abn & NGC~5194  & N/A & ILRT \\ 
AT2019krl & NGC~628  & 1,2 & ILRT \\ 
\bottomrule
\end{tabular}
\begin{flushleft}
\textit{Notes}: List of SNe and ILRT that occured during the survey. SNe that satisfied the `criteria' described in Section~\ref{sec:can_sel} are labelled as such.
\end{flushleft}
\end{center}
\end{table}

SNe are discovered in our survey due to the luminosity changes of the progenitor (satisfying criterion 1) or due to the transient SN itself (satisfying criterion 2). Table~\ref{tab:sne} presents the SNe that have occured during our survey and their classifications.  We also list the candidate criteria described in Section~\ref{sec:can_sel} that the SNe satisfied to be considered by our candidate detection pipeline.  Most of these SNe were discovered as luminous transients, satisfying criterion 2.  SN~2008S was first observed while in outburst and was thus not flagged as a transient. The fading of the SN was observed as a rapidly fading ``progenitor'', thus satisfying criterion 1.  

As a Type Ia, SN~2011fe has effectively no progenitor luminosity, and since the SN itself has faded significantly, this source does not satisfy criterion 1.  SN~2009hd, a Type II-L in NGC~3627, and SN~2012fh, a Type Ic in NGC~3344, satisfy neither criteria due to the relative faintness of the progenitors and the SNe in our data. SN~2009hd was heavily extincted due to the progenitor being in a dust lane of the galaxy, and a more-detailed analysis of SN~2012fh can be found in G15, A17b, and \citet{johnson17}.  SN~2013am and SN~2013ej are still more luminous than their progenitors but with $\Delta \nu L_\nu <10^4~\lsun$, so they did not satisfy criterion 1.

Of the 6 SNe identified under criterion 1, it is worth commenting that 4 (SN~2014J, SN~2014bc, SN~2016cok, and SN~2017eaw) were flagged because the SNe are still significantly brighter than their progenitors at the end of our observations. SN~2014bc is suspect as it lies in a highly saturated region of the $R$-, $V$-, and $B$- band data, leaving only a rather noisy $U$-band detection. Only SN~2011dh is flagged because the SN is significantly fainter than the progenitor so that the system is flagged through the death of the star.   Eventually, we would expect this to be true for most of these other SNe, but we consider the question of how long it takes a SN to become fainter than its progenitor.

The late time luminosity of a Type II-P SN roughly follows the radioactive decay of ${}^{56}$Ni
$$ L(t) = 1.45\ee{43} \bigg(\frac{M_\text{Ni}}{\msun}\bigg) e^{-t/\tau_0} {\rm erg~s^{-1}}$$
where $M_\text{Ni}$ is the mass of ${}^{56}$Ni and $\tau_0 = 111.3\rm~d$ is the effective decay rate (e.g., \citealt{nadyozhin94}).  We can invert this to find the time for the source to fade below luminosity $L$ where we should be able to detect most of these stellar deaths by the fading of the progenitor once $L<10^4~\lsun$ 
$$ t_\text{fade} = 3.2 + 0.7 \log_{10} \bigg[ \bigg(\frac{10^4~\lsun}{L}\bigg) \bigg(\frac{M_\text{Ni}}{0.1~\msun}\bigg) \bigg]\text{ yr} $$
Given that Type II-P SNe produce $M_\text{Ni} < 0.1~\msun$ \citep{hamuy03,sukhbold16}, it follows that the timescale for fading is around 3~yr.  This ignores $\gamma$-ray escape at late times, which will accelerate the fading, and other radioactive elements like ${}^{57}$Co and ${}^{44}$Ti, which can produce some luminosity at late times, though usually $<10^4~\lsun$ \citep{seitenzahl14}.

The other possible source of luminosity is shocked material, which can produce significant luminosity even after the ${}^{56}$Ni decay.  The luminosity of a shock interacting with a spherically symmetric pre-SN mass-loss wind (see e.g., \citealt{chevalier82}) is
\begin{gather*}
 L_\text{s} \simeq \frac{1}{2}\dot{M}\frac{v_\text{s}^3}{v_\text{w}} \\ 
\simeq 8000 \bigg( \frac{\dot{M}}{10^{-6}~\msun {\rm ~yr^{-1}}} \bigg) \bigg( \frac{v_\text{s}}{10^3 {\rm ~km~s^{-1}}} \bigg)^3 \bigg( \frac{10{\rm ~km~s^{-1}}}{v_\text{w}} \bigg) ~\lsun
\end{gather*}
where $\dot{M}$ is the mass-loss rate, and $v_\text{s}$ and $v_\text{w}$ are the shock and wind speeds, respectively.  This means that the shocked material can produce significant luminosity with a high shock speed or a high mass-loss rate.  While this is most likely to be emitted as X-rays because of the high post-shock temperature \citep{chevalier82}, some of the energy emerges as optical line emission as seen in the late-time spectra of SN~1980K and SN~1993J (e.g., \citealt{milisavljevic12}).  While it does not seem an ideal mechanism for normal SN, shocked material could support the late-time luminosity of SN~2013am, SN~2013ej, and SN~2014bc, where luminosity from radioactivity is unlikely to power the light curves more than 5~yr after the initial SNe.  

We also flagged the intermediate luminosity red/optical transient (ILRT/ILOT) AT2019krl (ZTF19abehwhj).  Located at $(\alpha, \delta) =$ (01:36:49.65, +15:46:46.2) in NGC~628, the transient was detected as a bright flare satisfying criterion 2, and also satisfied criterion 1 because it has not yet faded.  The LBT data on this transient is discussed in \citet{andrews20}.  Another ILRT/ILOT, AT2019abn (ZTF19aadyppr), occured in NGC~5194 at $(\alpha, \delta) =$ (13:29:42.39 +47:11:17.0) over the observing period of our survey, but it was not detected or analyzed because it is bright and saturated in our last epoch of candidate selection.  

\section{Candidates}\label{sec:candidates}

\begin{table*}
\caption{Candidate List}
\label{tab:candidates}
\begin{tabularx}{0.85\textwidth}{lrrcccc} \toprule
	&	&	& Candidate & $L_{R,\text{i}}-L_{R,\text{f}}$ & $L_{R,\text{max}}-L_{R,\text{min}}$  & 		\\
\multicolumn{1}{c}{ID}	& \multicolumn{1}{c}{RA} & \multicolumn{1}{c}{Dec} & Criteria  & [$\lsun$]& [$\lsun$] & Classification \\
\midrule
N4736-OC1 & 12:51:00.93 & +41:08:30.4 & 1  & $9.7\times10^{3}$ & $\hphantom{>}1.2\times10^{4}$  & OC \\
M101-OC1 & 14:03:17.24 & +54:22:07.6 & 1     & $9.9\times10^{3}$ & $\hphantom{>}1.2\times10^{4}$ & OC \\ \midrule
N6946-BH1 & 20:35:27.56 & +60:08:08.3 & 1,2      & $6.0\times10^{4}$ & $\hphantom{>}7.7\times10^{5}$  & FSN \\
M101 OT2015-1 & 14:02:16.80 & +54:26:20.7 & 1,2   & $1.1\times10^{5}$ & >$1.8\times10^{6}$  & merger \\ 
SN~2011dh & 13:30:05.15 & +47:11:11.8 & 1,2  & $5.7\times10^{4}$ & >$6.5\times10^{6}$ & SN \\ 
\bottomrule
\end{tabularx}
\begin{flushleft} \textit{Notes}: List of candidates that passed the final round of visual inspection. `Candidate criteria' are those listed in Section~\ref{sec:can_sel}. $L_{R,\text{i}}$ and $L_{R,\text{f}}$ are the $R$-band luminosities of the first and last epochs, while $L_{R,\text{max}}$ and $L_{R,\text{min}}$ are the maximum and minimum $R$-band luminosities observed for each source in the LBT light curves. N6946-BH1 was previously identified as a failed SN, and M101~OT2015-1 and SN~2011dh are included as candidates due to the ``disappearence'' of their progenitor sources.  FSN = failed SN, OC = other candidate. \end{flushleft}
\end{table*}

Here we describe and analyze the remaining five candidates.  Of these, three are ``rediscovered'' candidates discussed in G15, A17a, A17b, and B20: N6946-BH1, M101~OT2015-1 (PSN J14021678+5426205), and SN~2011dh. N6946-BH1 satisfied both selection criteria in the \textit{BVR}-bands by producing a luminous optical transient and fading significantly below the progenitor luminosity (for details, see A17a).  M101~OT2015-1 was a luminous red nova and suspected stellar merger \citep{goranskij16,blagorodnova17}, and the associated LBT data are discussed in detail in A17b.  The source satisfied both candidate criteria by producing a luminous transient and fading below its progenitor luminosity.  Considering the LBT data alone, without outside analysis/identification, would have lead us to classify this source, along with SN~2011dh as discussed earlier, as a failed SN, hence we include these in Table~\ref{tab:candidates}.  These sources served as benchmarks for our analysis - that we were able to recover them means that our methods are consistent and we would likely not miss similar sources.

We consider two new candidates, N4736-OC1 and M101-OC1, where a luminous source is present in many or most of the epochs of observation before fading significantly.  Neither source is as strong a failed SN candidate as N6946-BH1. Both sources are less luminous and far bluer than N6946-BH1 and were not observed to produce a luminous optical transient. Based on new data from January and March 2021 that was not used for candidate selection, we ultimately reject N4736-OC1 as a candidate failed SN.

\subsection{N4736-OC1}\label{sec:n4736}

\begin{figure*}
\centering
\includegraphics[width=\linewidth]{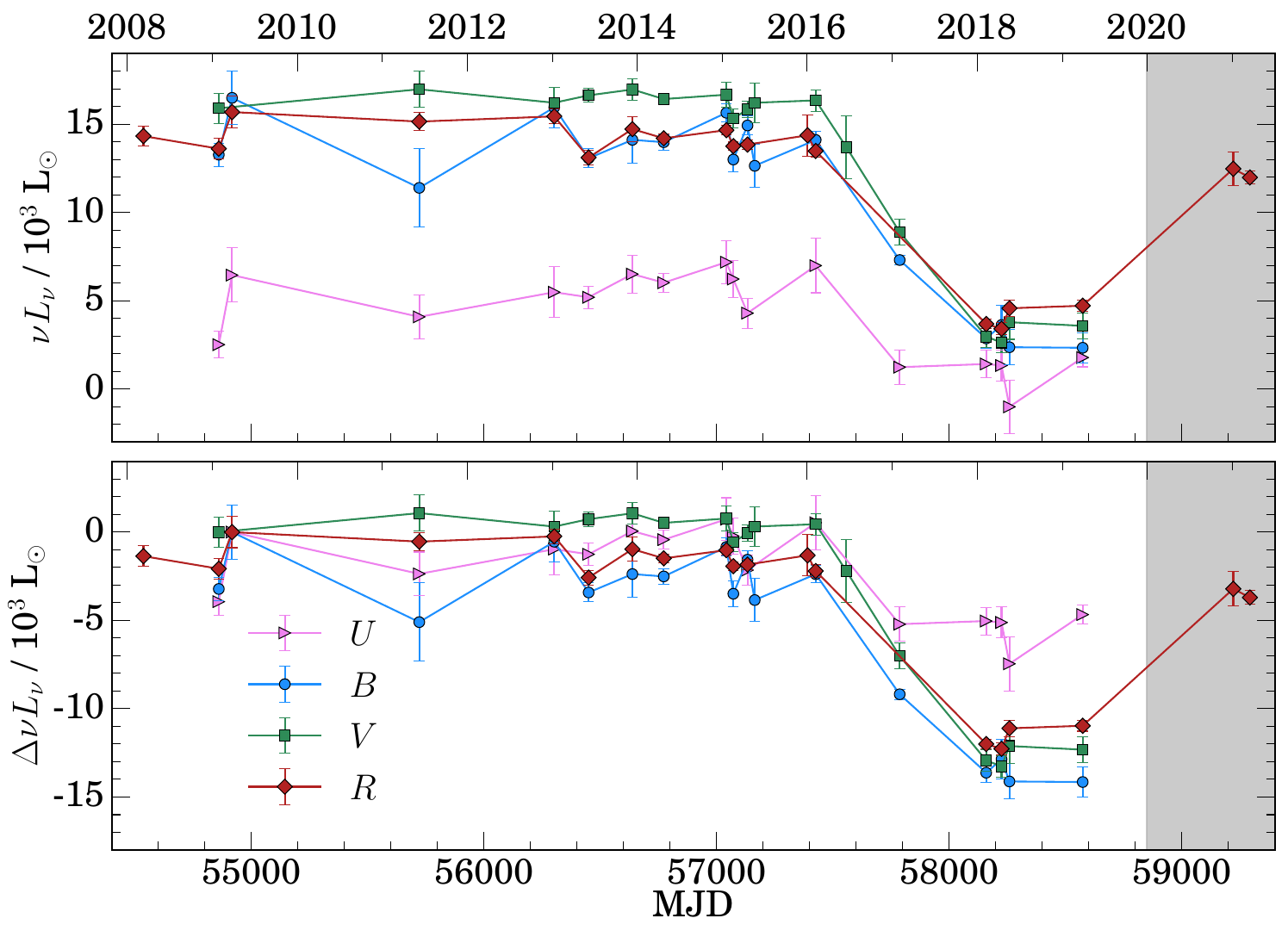}
\caption{\textbf{Top:} Calibrated light curve of N4736-OC1. \textbf{Bottom:} Differential light curve of N4736-OC1.  Luminosity is relative to the second epoch of 2009 for all bands except $V$-band, which is relative to the first epoch of 2009. The dark-grey shaded region are epochs that were not included in candidate selection.} 
\label{fig:n4736-lc}
\end{figure*}

\begin{figure*}
\centering
\includegraphics[width=\linewidth]{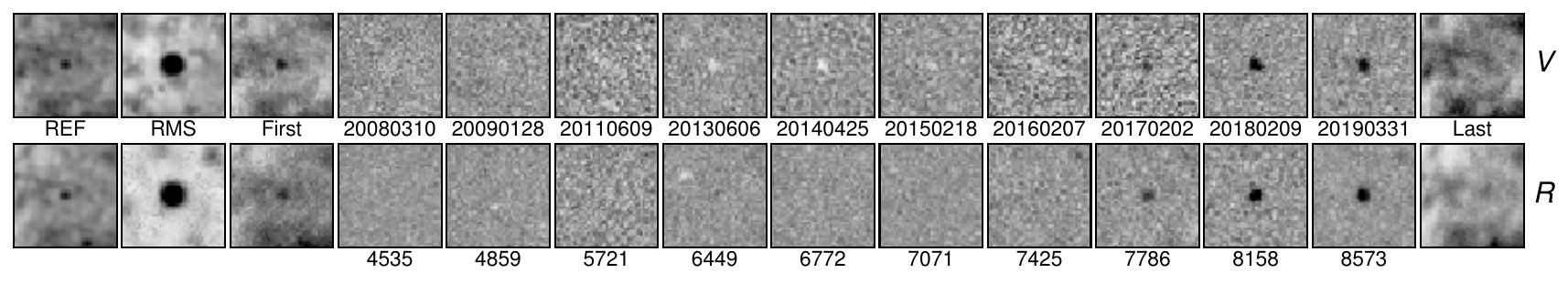}
\caption{$V$- and $R$-band imaging of N4736-OC1.  In the reference (REF), RMS, First and Last images, the darker the source, the brighter it is.  First refers to the first image where the seeing is good enough to clearly identify the source, not necessarily the earliest image in the survey, whereas Last is the most recent image that was included in candidate selection.  The middle images labeled with dates (top) and the last four digits of the JD (bottom) are the subtracted images, where the lighter shades mean the source is brighter than the reference image and darker shades mean the source is fainter than the reference image.  The two lines marking the same position in each image are both 4 pix = 0\farcs9.} 
\label{fig:n4736-sub}
\end{figure*}

N4736-OC1 is in NGC~4736 at $(\alpha,\delta)=$ (12:51:00.93, +41:08:30.4).  The calibrated and differential light curve of the source are shown in Figure \ref{fig:n4736-lc}, and the LBT subtraction images are shown in Figure \ref{fig:n4736-sub}.  The full light-curve data are shown in Table \ref{tab:n4736}. The source remained relatively quiescent for the first 8~yr of the survey.  The source then fades by $\sim$10$^4~\lsun$ over 2~yr in the \textit{BVR}-bands and by $\sim$5$\times10^3~\lsun$ in the $U$-band.  While there is still some residual flux in the final epochs of LBT data (see the top panel of Fig.~\ref{fig:n4736-lc}), the point source that is present in earlier epochs is clearly absent in the later epochs (see Fig.~\ref{fig:n4736-sub}).  The source was not flagged as a candidate in the $R$- or $U$-bands because the $\Delta \nu L_\nu$ measured for criterion 1 was slightly less than $10^4~\lsun$ in both bands.  However, it did satisfy criterion 1 in $B$- and $V$-bands.   

There are unfortunately no \hst~images of the source, limiting our ability to confirm or reject it as a true candidate. There is no evidence for a source or dust emission in the available \spitzer~3.6 and 4.5 micron images from MJD 56906 (September 2014) to 58778 (October 2019).  Light curves extracted at the location of the source in \spitzer~images have an RMS scatter of $\sim$10$^4~\lsun$, slightly below the level of the optical luminosity changes, but with no evidence of systematic flux changes between the earlier and later epochs.  While we cannot rule out dust converting some of the optical flux into mid-IR flux, it seems unlikely that dust is the driving mechanism for the optical luminosity change.

Based on the differential light curve, the progenitor star was likely $\sim$10$^4~\lsun$, which is quite under-luminous for a supergiant.  One explanation is that the source is a post-AGB star \citep{bloecker95}.  Such stars have been known to fade significantly in optical wavelengths due to the creation of dust shells (e.g., V4334~Sgr/Sakurai's object \citealt{duerbeck00}).  Another alternative is that the source is an R Coronae Borealis (RCB) star.  RCBs are cool supergiants that can fade several magnitudes for up to thousands of days due to variable dust creation \citep{clayton12}.  RCBs also have absolute magnitudes of $M_V \sim -2.6~\text{to}~-5.2$ \citep{tisserand09}, translating to roughly $\nu L_\nu \sim 10^{3}$$-$$10^{4}~\lsun$, which is comparable to what we see in Figure \ref{fig:n4736-lc}.   Both of these explanations may be problematic due to the lack of mid-IR flux increase.  

We obtained new LBT data in January and March 2021 and found that the star has returned to near peak brightness (see Fig.~\ref{fig:n4736-lc}).  Thus, we ultimately reject N4736-OC1 as a failed SN candidate. We include this analysis of the source because, considering only the data used in candidate selection, it was a viable candidate, and it illustrates the risk that hitherto-unknown stellar variability could very well mimic a failed SN.

\subsection{M101-OC1}\label{sec:m101}

\begin{figure*}
\centering
\includegraphics[width=\linewidth]{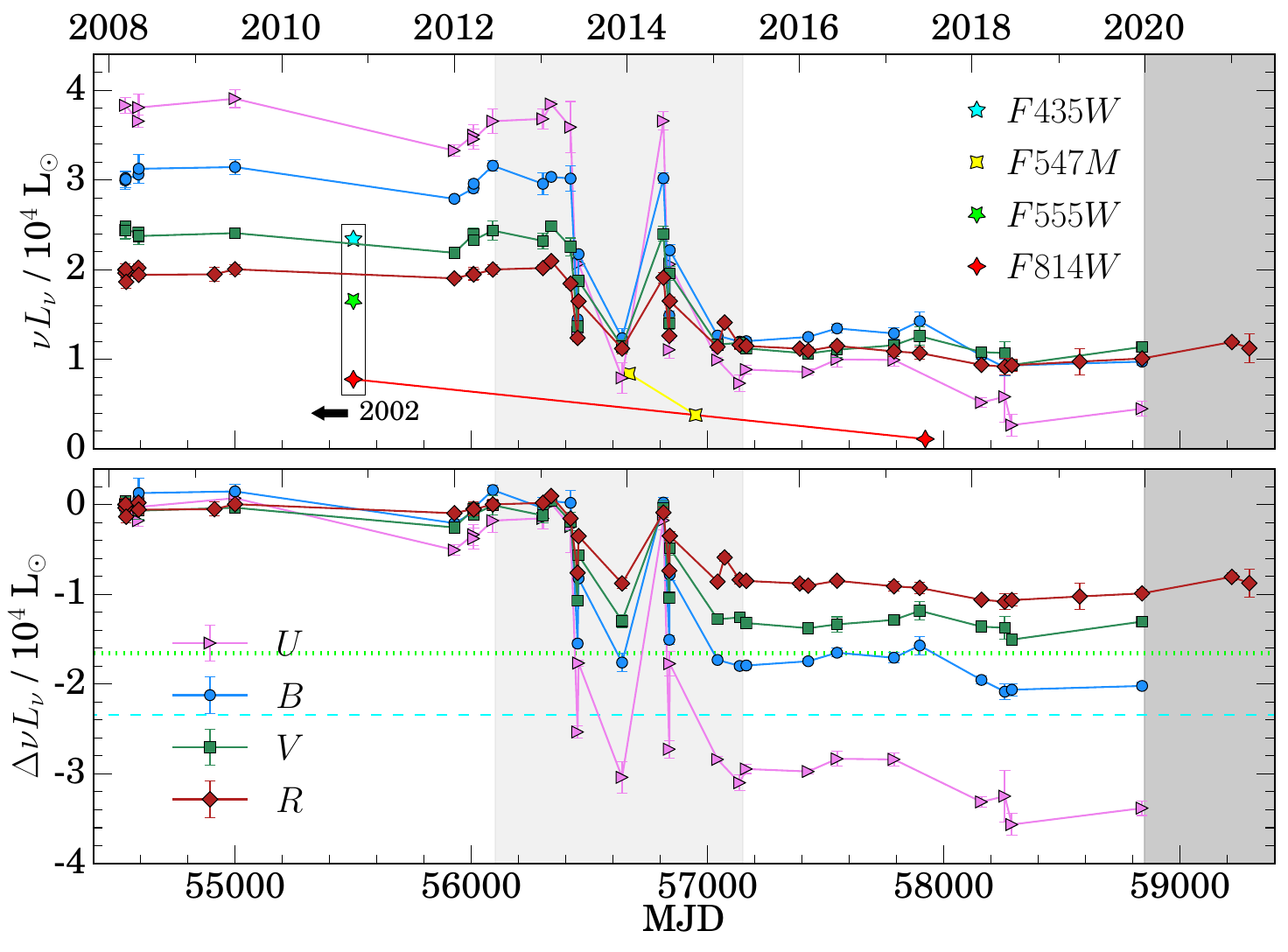}
\caption{\textbf{Top:} Calibrated light curve of M101-OC1, including the \hst~photometry.  The \hst~photometry that are boxed are from 2002 but are shifted in time so as to fit in this plot. \textbf{Bottom:} Differential light curve of M101-OC1.  Luminosity is relative to the first epoch of observation. The range of dates highlighted in light-grey are those focused on in Figure \ref{fig:m101-lczoom}.  The cyan/dashed and lime/dotted lines are the negative $F435W$ and $F555W$ fluxes, meant to highlight that the disappearing $B$ and $V$ flux seen by LBT very nearly matches the $F435W$ and $F555W$ flux seen by \hst. The dark-grey shaded region are epochs that were not included in candidate selection.} 
\label{fig:m101-lc}
\end{figure*}

\begin{figure*}
\centering
\includegraphics[width=\linewidth]{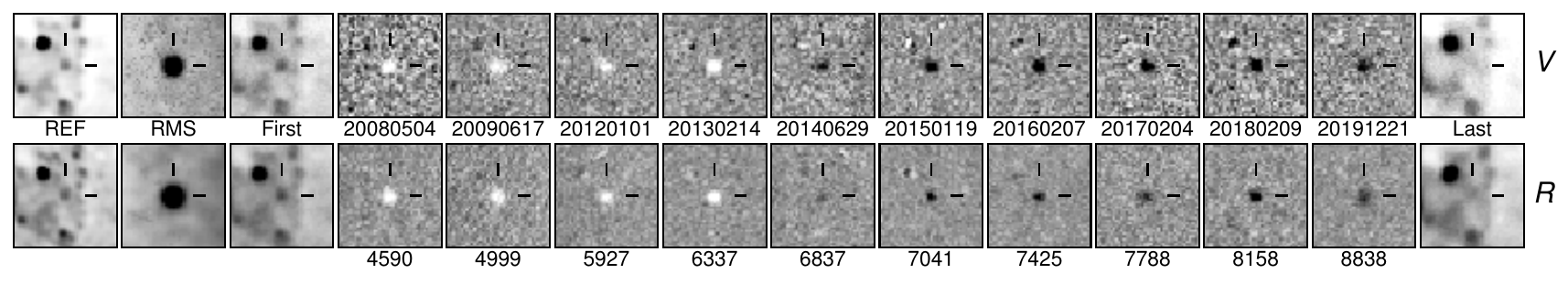}
\caption{Same as Fig. \ref{fig:n4736-sub}, but for M101-OC1.}
\label{fig:m101-sub}
\end{figure*}

M101-OC1 is in M101 at $(\alpha,\delta)=$ (14:03:17.24, +54:22:07.6).  The calibrated and differential light curves of the source are shown in Figure \ref{fig:m101-lc}, and the LBT difference imaging is shown in Figure \ref{fig:m101-sub}.  The full light-curve data are shown in Table \ref{tab:m101}.  The source remained quiescent for the first 5~yr of observations, before showing peculiar variability for $\sim$2~yr between early 2013 and late 2014.  After this ``episode'' of variability, the source dropped in flux by $10^4~\lsun$ in the $R$-band.  This drop was more extreme in the bluer bands, with $\Delta \nu L_\nu \sim 3\times10^4~\lsun$ in the $U$-band (see Fig.~\ref{fig:m101-lc}).  The point source present in the early LBT \textit{UBV} epochs is not visible in the latest epochs, though there appears to be a faint source in the last few $R$-band images (see Fig.~\ref{fig:m101-sub}).  Similar to N4736-OC1, the sourced was flagged in the \textit{UBV}-bands for satisfying criterion 1 but was not flagged in the $R$-band.  We obtained new LBT data in January and March 2021 which show the source to have remained relatively quiescent and faint (see Fig.~\ref{fig:m101-lc}).

\begin{figure}
\centering
\includegraphics[width=\linewidth]{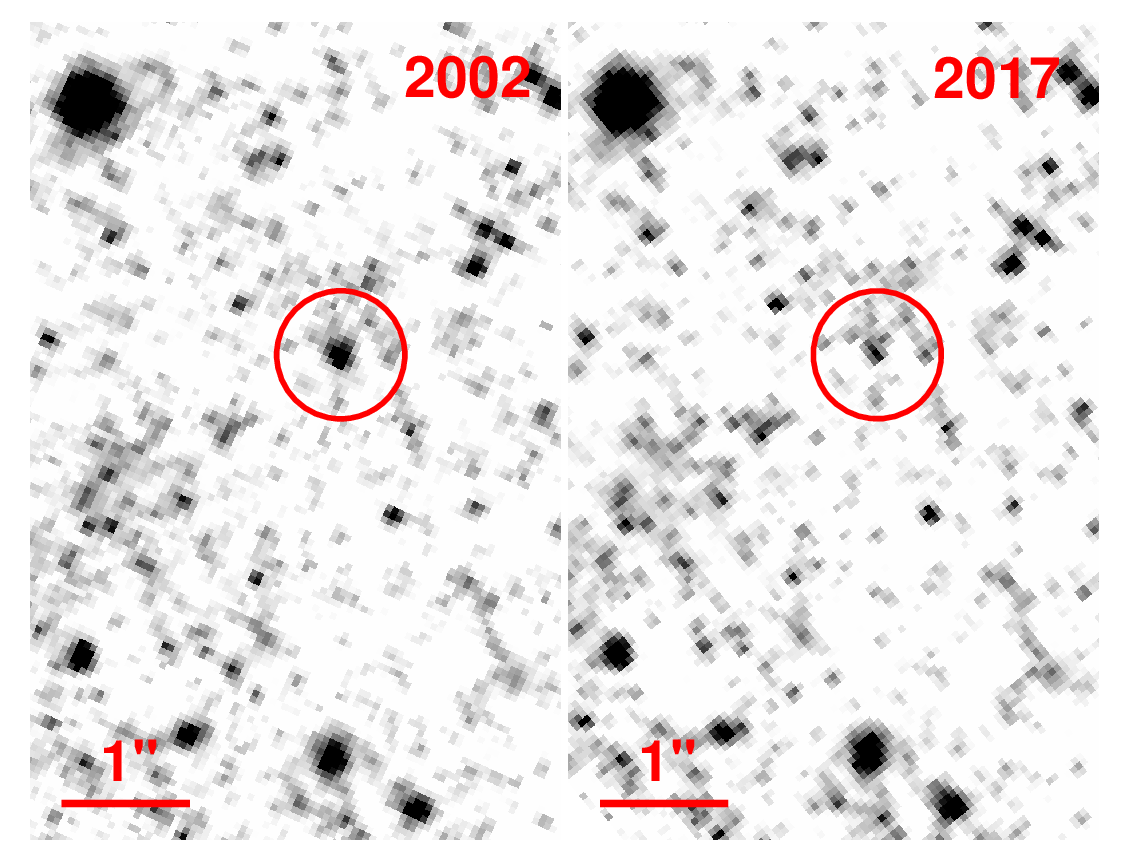}
\caption{$F814W$~\hst~images of M101-OC1 (circled) from 2002 and 2017.} 
\label{fig:m101-hst}
\end{figure}

\begin{table*}
\caption{\hst~photometry}
\label{tab:hst}
\begin{center}
\begin{tabular}{lcccccl} \toprule
 \multicolumn{1}{c}{Filter} & Date & Epoch (MJD) & $m$ (mag) & $M$ (mag) & $\nu L_\nu$ ($\lsun$) & \multicolumn{1}{c}{Reference} \\ \midrule
$F435W$ & \multirow{2}{*}{2002-11-15} & \multirow{2}{*}{52594} & 22.998 $\pm$ 0.014 & $-$6.072 & 23413 & \multirow{2}{*}{PI: K. Kuntz, GO-9490} \\
$F555W$ & & & 23.036 $\pm$ 0.016 & $-$6.037 & 16548 & \\ \midrule
\multirow{2}{*}{$F814W$} & 2002-11-15 & 52594 & 22.958 $\pm$ 0.025 & $-$6.093 & 7778 & PI: K. Kuntz, GO-9490 \\
 & 2017-06-17 & 57921 & 24.433 $\pm$ 0.019 & $-$4.619 & 1999 & PI: B. Shappee, GO-14678\\ \midrule
\multirow{2}{*}{$F547M$} & 2014-01-10 & 56667 & 23.742 $\pm$ 0.041 & $-$5.321 & 8405 & \multirow{2}{*}{PI: W. Blair, GO-13361} \\
 & 2014-10-18 & 56949 & 24.605 $\pm$ 0.101  & $-$4.458 & 3796 &  \\
\bottomrule
\end{tabular}
\begin{flushleft}
\textit{Notes}: $M$ and $\nu L_\nu$ are corrected for Galactic extinction, $m$ is not.  See Tab.~\ref{tab:sample} for distance modulus and reference.
\end{flushleft}
\end{center}
\end{table*}

Serendipitously, there were multiple \hst~observations of M101-OC1 across different epochs and with different filters.  The resultant photometry was extracted with \texttt{DOLPHOT} \citep{dolphin00} using the same configuration described in \citet{adams15} and using the drizzled $F814W$ image from 2002 as the reference for the source's location.  

The \hst~PSF photometry magnitudes are presented in Figure \ref{fig:m101-lc} and Table \ref{tab:hst}.  $F814W$ is the only filter used to image the source both before and after the fading incident.  The pre- and post-fading $F814W$ images are shown in Figure \ref{fig:m101-hst}.  While the source is significantly brighter in the pre-fading image from 2002, there is still a clear point source in the post-fading image from 2017.  There are also two epochs with the $F547M$ filter that show the source fading over time, though they occur during the episode of variability, and so it is unclear how these data relate to the other LBT and \hst~data. The lower luminosities found for the \hst~filters (e.g., comparing $F435W$ with $B$, $F555W$ with $V$) are likely due to crowding in the lower resolution LBT data.  It is worth noting that the \hst~luminosities are comparable to the luminosity changes measured in the LBT data (see the bottom panel of Fig.~\ref{fig:m101-lc}), which is expected given that differential light curves are insensitive to crowding.  Future \hst~observations, especially at bluer bands like $F435W$ and $F555W$ will be needed to better understand the source.

Based on differencing the 3.6 and 4.5 micron \spitzer~images, there is a source with some dust emission at the position of the star, but it is too confused to identify in the individual images.  It also shows no evidence of a change in mid-IR luminosity that is on the scale of the change in the optical emission, to limits of $\leq 2 \times 10^4~\lsun$ based on RMS variability, over the period of MJD 53072 (March 2004) to 58781 (October 2019). This implies that the intrinsic luminosity of M101-OC1 is dropping rather than being shifted from optical to mid-IR wavelengths. Although the missing luminosity could be hidden in emission from colder dust that would not be detected by \spitzer~in the mid-IR, one might expect a dust formation phase with hot dust which would have been detectable. 

From the significant drop in luminosity in the blue bands, M101-OC1 is a candidate for a 10$^4$--10$^5~\lsun$ blue supergiant (BSG) that disappeared as a failed SN.  As a BSG, a failed SN would not be expected to produce a significant transient like N6946-BH1, as BSGs are more compact than YSGs/RSGs.  After core collapse, a small fraction of the BSG's envelope would be ejected and recombine, producing only a short-lived ($\sim$20~d) transient \citep{fernandez18}.  Because the galaxies in the survey are observed a few times a year, it is unlikely that we would observe such a short-lived event.  While there is still a faint point source in the late-time LBT and \hst~observations, this could be understood as emission from fallback accretion or a faint red binary companion to the progenitor.  The latter explanation might be favored, as BSGs are very likely to have binary companions \citep{sana12}.  Furthermore, the unchanging flux from dust emission seen by \spitzer~could be attributed to this hypothetical red companion, since hot stars cannot form dust \citep{kochanek11}.

If M101-OC1 is not a failed SN and is instead a strange variable star, one possible interpretation is that it is a dust-obscured LBV.  LBVs undergo ``outbursts'' where they become significantly cooler while maintaining their intrinsic luminosities \citep{humphreys94}, and could thus mimic a disappearance in optical wavelengths by becoming sufficiently cool.  LBVs are typically much more luminous at $\sim$10$^6~\lsun$, but if there is significant obscuration by cold dust, then the optical luminosity could perhaps be brought down to the level of 10$^4$--10$^5~\lsun$ that we see before M101-OC1 disappears, although this would make the relatively blue color of the optical source difficult to explain.  The disappearance of M101-OC1 could then be attributed to conventional LBV behavior.  This explanation is problematic, as M101-OC1 does not appear to maintain its intrinsic luminosity as it fades, as evidenced by the lack of change in the mid-IR \spitzer~flux, whereas LBVs are thought to vary at roughly fixed luminsity except in outburst \citep{humphreys94}. 

If we consider other options, M101-OC1 could be a hot ($T > 15000$~K) RCB star undergoing a dimming episode.  However, issues immediately arise with this idea. There are only five hot RCBs known and even fewer with reported distances \citep{demarco02,tisserand20}, and M101-OC1 is also significantly more luminous than any hot RCB with a known distance. 

Finally, M101-OC1 could be understood as a BSG being enshrouded in cold dust.  There is a \spitzer~source that is likely due to some warm dust emission, and so some of this warm dust may have cooled and enshrouded the star, leading to a drop in optical flux. However, this scenario is somewhat contrived, and it does not explain how the warm dust is being created, since the BSG is unlikely to be producing significant dust by itself \citep{kochanek11}.  Without more data, we are unable to say more.  The \textit{James Webb Space Telescope} (\textit{JWST}) is needed to determine the importance of cold dust in understanding this source.

\begin{figure}
\centering
\includegraphics[width=\linewidth]{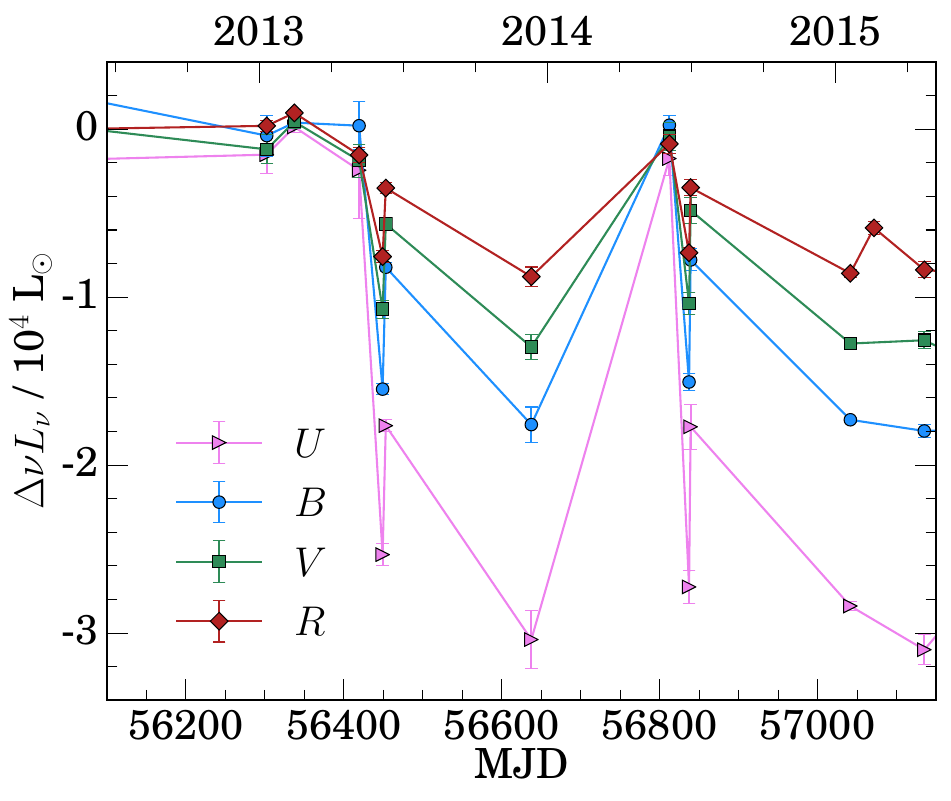}
\caption{Zoomed-in differential light curve of M101-OC1.  Notice the strange variability exhibited between late-2013 and mid-2014.  After this, the source faded and stayed significantly less luminous than the progenitor luminosity.}
\label{fig:m101-lczoom}
\end{figure}

\begin{figure*}
\centering
\includegraphics[width=\linewidth]{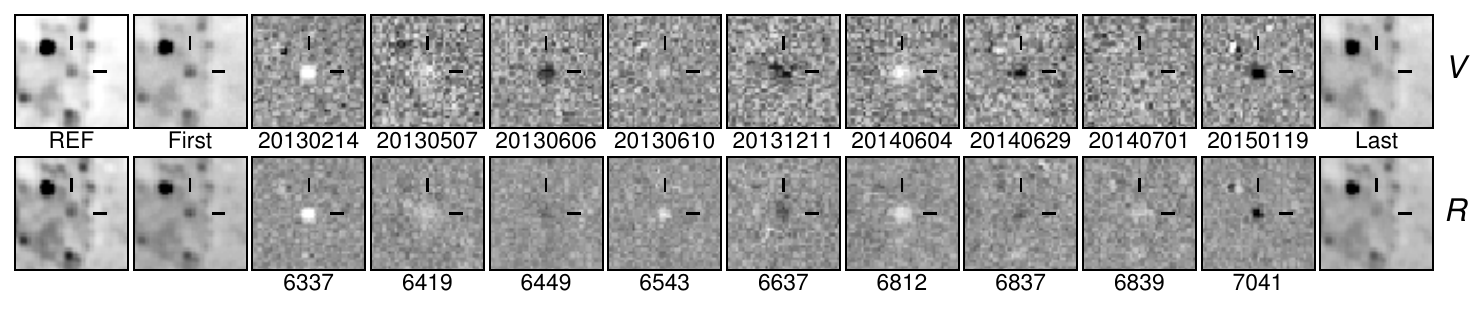} 
\caption{Same as Fig. \ref{fig:m101-sub}, but focusing on the epochs during the episode of variability.  Here, ``First'' and ``Last'' refer to the epochs before and after the variability.} 
\label{fig:m101-zoom}
\end{figure*}

The ``episode'' of variability prior to the source's fading is very peculiar.  Figure \ref{fig:m101-lczoom} shows a zoom-in of the episode of variability in the differential light curve, and Figure \ref{fig:m101-zoom} shows the relevant LBT subtraction images.  Nearby sources do not show any similar variability, and the variability is replicated in all four bands.  This is important, as the \textit{UBV}-band images are from a different primary mirror and camera on LBT than the $R$-band images.  This makes it extremely likely that this variability is real and not an artefact of the data, flux calibration, or image subtraction.

The last epoch before the episode of variability is MJD~56337 (February 2013). Following this epoch, the source undergoes a roughly month-long rise to a peak luminosity, rapidly dimming, and then returning to near peak luminosity.  It does this twice, during MJD~56419$-$56453 (May-June 2013) and MJD~56812$-$56839 (June-July 2014).  In the middle of these two trends, MJD~56638 (December 2013) the source is observed to have faded significantly.   The episode of variability appears to end by the next epoch of MJD~57042 (January 2015), where the source has faded and remains at or below this luminosity for the rest of our observations.  The episode of variability lasts 1--2~yr. \spitzer~data do not show the source to undergo similar variability in the mid-IR bands, though the closest match in time is the 3.6 micron \spitzer~epoch taken 25~d prior (MJD~56394) to the first epoch of variability.  Both of the two epochs of \hst~data in $F547M$ were observed during the episode of variability, though it is unclear how either epoch fit in with the variability seen with LBT due to the mismatch between the \hst~and LBT fluxes (see earlier discussion of \hst~photometry).  

It is unclear what is producing this variability.  One possible interpretation is variable dust creation, which could cover and then expose the progenitor BSG repeatedly.  This could explain why the source is able to return to near-progenitor luminosity in the later part of the episode.  While \spitzer~data do not show any mid-IR variability that could confirm this, the rapid timescales for evolution (i.e., the source brightening by $\sim$10$^4~\lsun$ in 2~d) and the mismatch between \spitzer~and LBT epochs could explain why we do not see \spitzer~variability.  As noted earlier, this variable dust creation scenario is unlikely since BSGs are not known to produce significant quantities of dust.  Another possibility is that we are observing eclipses of the BSG by a binary companion, which fits in with the remnant optical flux seen in LBT and \hst~images and the roughly constant mid-IR source seen in \spitzer.  From the three epochs during the episode of variability where the source is faintest, we derive an approximate period of $\sim$194~d.  When we phase-fold the light curve by this period, we find that none of the earlier data points (when the source is consistently bright) land in the phase-space of the eclipse, meaning that the earlier data do not contradict this eclipse interpretation.  However, with so little data, we cannot definitively say that these were eclipses.

\subsection{Comparing new candidates to N6946-BH1}

\begin{figure*}
\centering
\includegraphics[width=\linewidth]{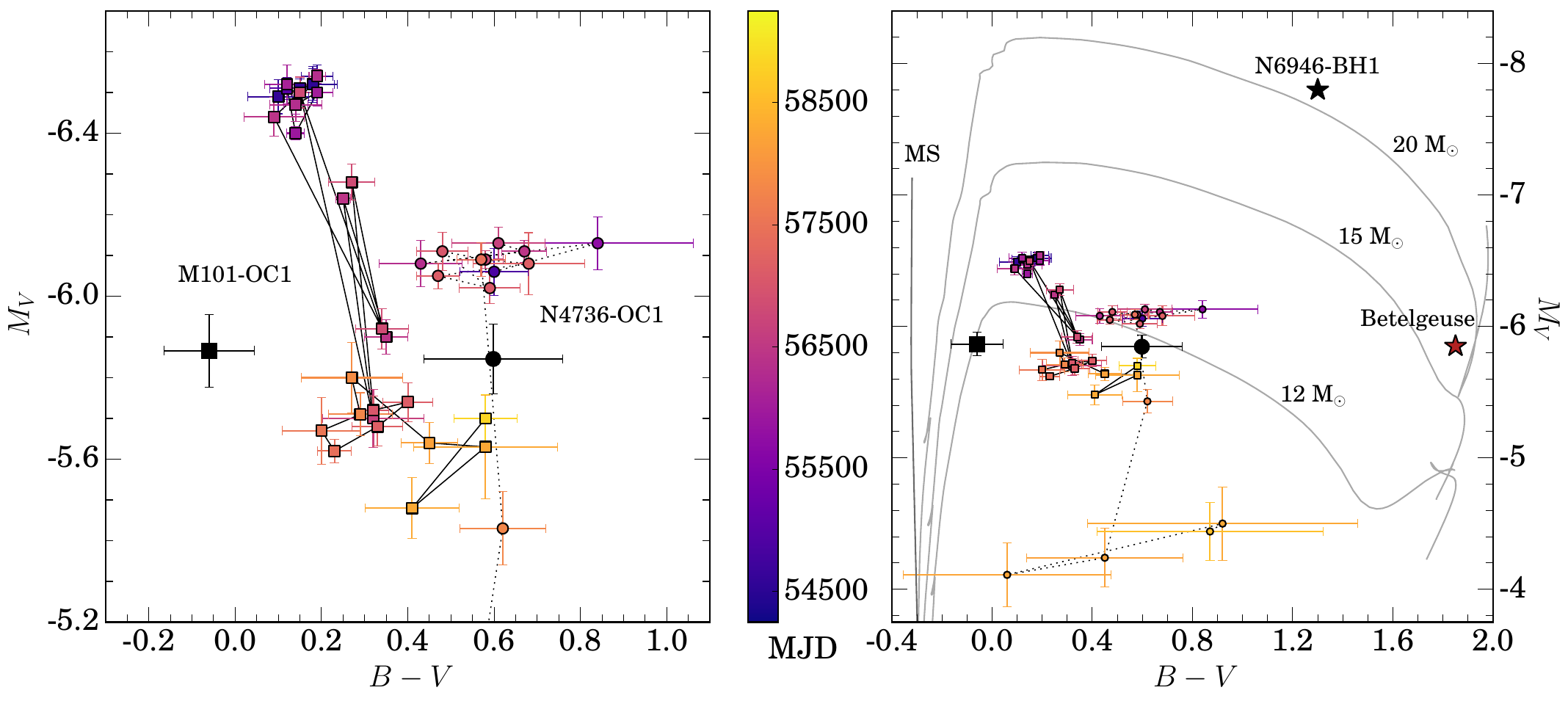}
\caption{\textbf{Left:} CMD of the calibrated light curves of N4736-OC1 (circles) and M101-OC1 (squares).  The colors of points correspond to the epochs of the data.  The black circle and square represent the mean difference in flux between early and late time data (i.e., the flux of the disappearing star) for N4736-OC1 and M101-OC1, respectively.   \textbf{Right:} Same as left, but with a larger range of color and magnitude.  We show MIST \citep{dotter16,choi16} stellar evolution tracks and a young ($\sim$300~kyr) main sequence (MS), all with solar metallicity, for comparison.  We also show the median progenitor data for N6946-OC1 from A17b and C20.  We also include Betelgeuse as a classic example of a RSG.}
\label{fig:can-cmd}
\end{figure*}

We show the evolution of N4736-OC1 and M101-OC1 in a color-magnitude diagram (CMD) in Figure~\ref{fig:can-cmd}.  Since thse are based on estimates of the total flux, they are sensitive to issues like crowding.  We also show estimates of the location in the CMD for the ``disappearing star'' of each source, shown as a black circle and square for N4736-OC1 and M101-OC1, respectively, based on the differential light curves.  While the ``disappearing star'' of N4736-OC1 is nearly 2~mag more luminous than the late-time flux, the ``disappearing star'' of M101-OC1 is only $\sim$0.5~mag more luminous than the late-time flux, though it is significantly bluer.  As proposed earlier in the paper, this late time flux could be interpreted as a redder binary companion or fallback accretion and also due to crowding caused by LBT's resolution limit. 

Overall, we can see that both of these sources are significantly bluer and fainter than N6946-BH1 and other RSGs.  If we compare these ``disappearing stars'' to MIST \citep{dotter16,choi16} stellar evolution tracks, both appear to lie along or near the track of a $\sim$12$~\msun$ star, which is less massive and less luminous than the 18--25$~\msun$ RSGs that are most commonly expected to become failed SNe.  However, MIST tracks are constructed assuming a single star with no binary interactions, and, since there is at least some evidence for M101-OC1 having a companion, these single star evolution models may not be appropriate.  Future analysis with binary evolution models (e.g., BPASS, \citealt{eldridge17}) may be possible with new data to constrain the remnant luminosity and spectral energy distribution.  

\section{Peculiar Stellar Variables}\label{sec:weird}

\begin{figure*}
\centering
\includegraphics[width=\linewidth]{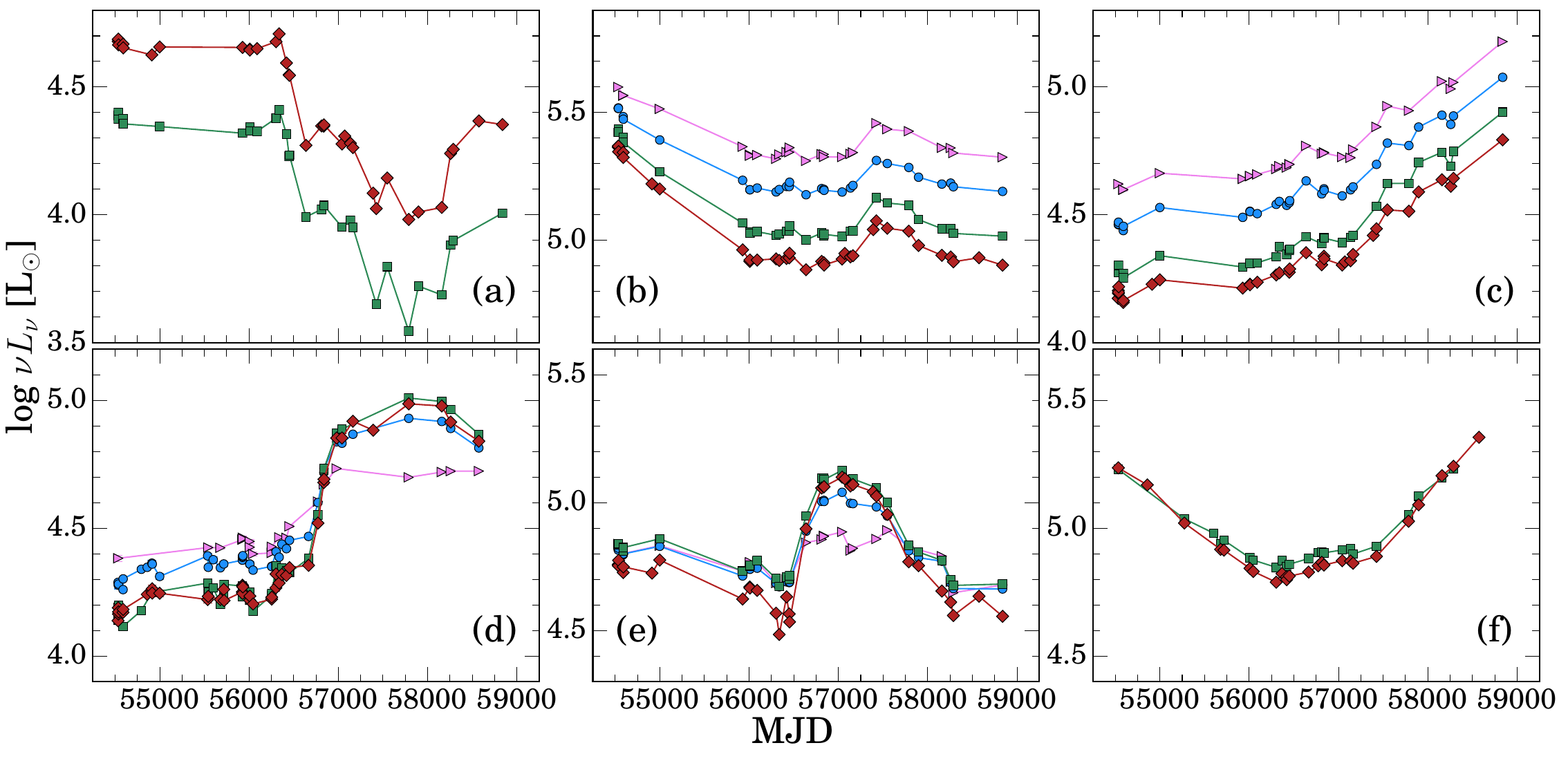}
\caption{Calibrated light curves of long-term, high-amplitude peculiar variable sources in our data set.  The luminosities vary between panels, but the dynamic range is constant at 1.3 dex.  Errors are based on \texttt{ISIS} errors, which are underestimates.  The $B$- and $U$-band data for sources (a) and (f) are problematic and so are not included. }
\label{fig:weird-lc}
\end{figure*}

\begin{figure*}
\centering
\includegraphics[width=\linewidth]{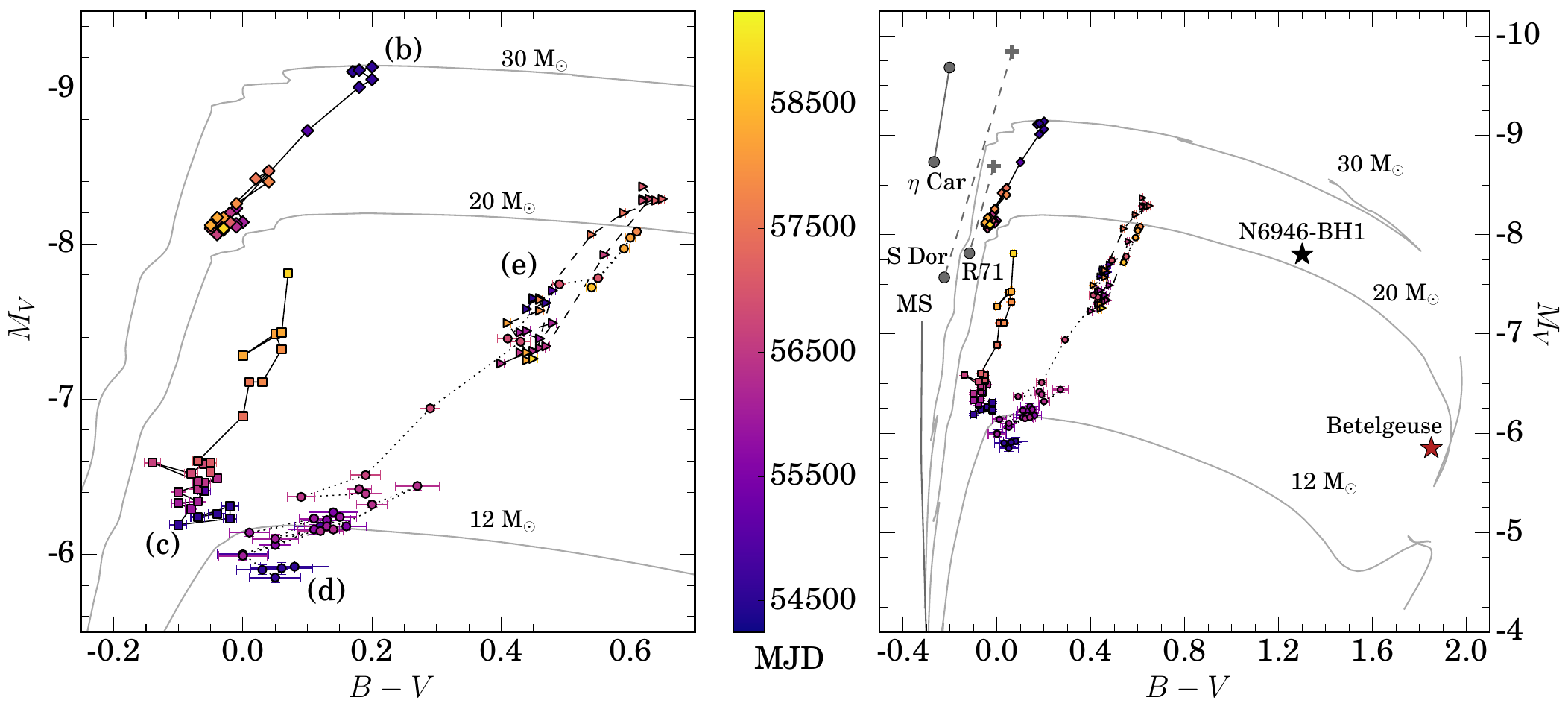}
\caption{Same as Fig.~\ref{fig:can-cmd}, but for the peculiar variable sources from Fig.~\ref{fig:weird-lc}.  Each source is represented by a different shape and is labelled as in Fig.~\ref{fig:weird-lc}.  The LBVs are converted to $B-V$ and $M_V$ from temperatures and luminosities listed in \citet{humphreys94}.  For the LBVs, circles represent ``quiescence'', and crosses represent ``eruptions''.}
\label{fig:weird-cmd}
\end{figure*}

In our search for failed SNe, we examine the light curves of many luminous stars with peculiar long-term variability.  In Figure \ref{fig:weird-lc}, we show a sample of these variable sources.  These sources slowly fade, brighten, or both, over the course of years. Some like source (c) change monotonically for over a decade, and some like source (f) exhibit more cyclic behavior.  The amplitudes of these luminosity changes are large, ranging from factors of 3 to 10, often with little change in the color.  This is significant, since LBVs usually show large changes in color.

We show the evolution of these sources in the CMD in Figure \ref{fig:weird-cmd} to better illustrate the changes in color.  We also show the approximate locations in the CMD of well-studied LBVs in quiescence and in eruption \citep{humphreys94}.  Source (b) closely mirrors known LBVs S~Doradus and R71 in evolution and in location on the CMD while sources (c), (d), and (e) appear too red and too faint to be LBVs.  Because we ultimately reject N4736-OC1 as a failed SN, N4736-OC1 could be considered among these peculiar variables. For now, we simply note these stars as a potentially new and interesting type of variable.  

\section{Conclusions}\label{sec:conclusion}
We update the LBT search for failed SN survey using a baseline of 11~yr of data.  We find:
\begin{itemize}
\item[$\bullet$] Our analysis re-discovers the original failed SN candidate, N6946-BH1.  We find two new ``disappearing stars'', N4736-OC1 and M101-OC1, both of which are bluer and less luminous than N6946-BH1 and were not observed to produce a transient flare. While neither of our new candidates resemble N6946-BH1, these are both interesting and peculiar objects. 
\item[$\bullet$] N4736-OC1 is possibly a post-AGB star. New LBT data from January and March 2021 show the source to have returned to near-peak brightness, and thus we reject the source as a candidate failed SN.
\item[$\bullet$] M101-OC1 is a very complicated source that is either a LBV that has become redder and less luminous by some unknown mechanism or a failed SN candidate with a fainter and redder binary companion. The evidence for the latter interpretation is the remnant optical flux seen by LBT and \hst, the non-transient mid-IR flux seen by \spitzer, and the possible eclipses in the LBT light curve.  While we favor this failed SN interpretation, more data from the ongoing LBT survey, \hst, and \textit{JWST} is needed confirm or reject M101-OC1 being a failed SN.
\item[$\bullet$] We present a small sample of peculiar, high-amplitude ($\Delta L/L > 3$) long-timescale ($\sim$decade) variable stars.  These stars appear to represent some previously unrecognized variable class which requires further exploration. 
\end{itemize}

Finally, we can update the estimates from G15 and A17b for the fraction of core collapses which fail to produce SNe. The failed SN/core-collapse fraction, $f$, is described by a binomial probability distribution function (PDF)
$$ P(f) \propto (1-f)^{N_\text{SN}} f^{N_\text{FSN}} $$
where $N_\text{SN}$ and $N_\text{FSN}$ are the number of successful and failed ccSNe, respectively.  In A17b, $N_\text{SN} = 6$, but now, with the addition of SN~2016cok and SN~2017eaw, $N_\text{SN} = 8$.

\begin{figure}
\centering
\includegraphics[width=\linewidth]{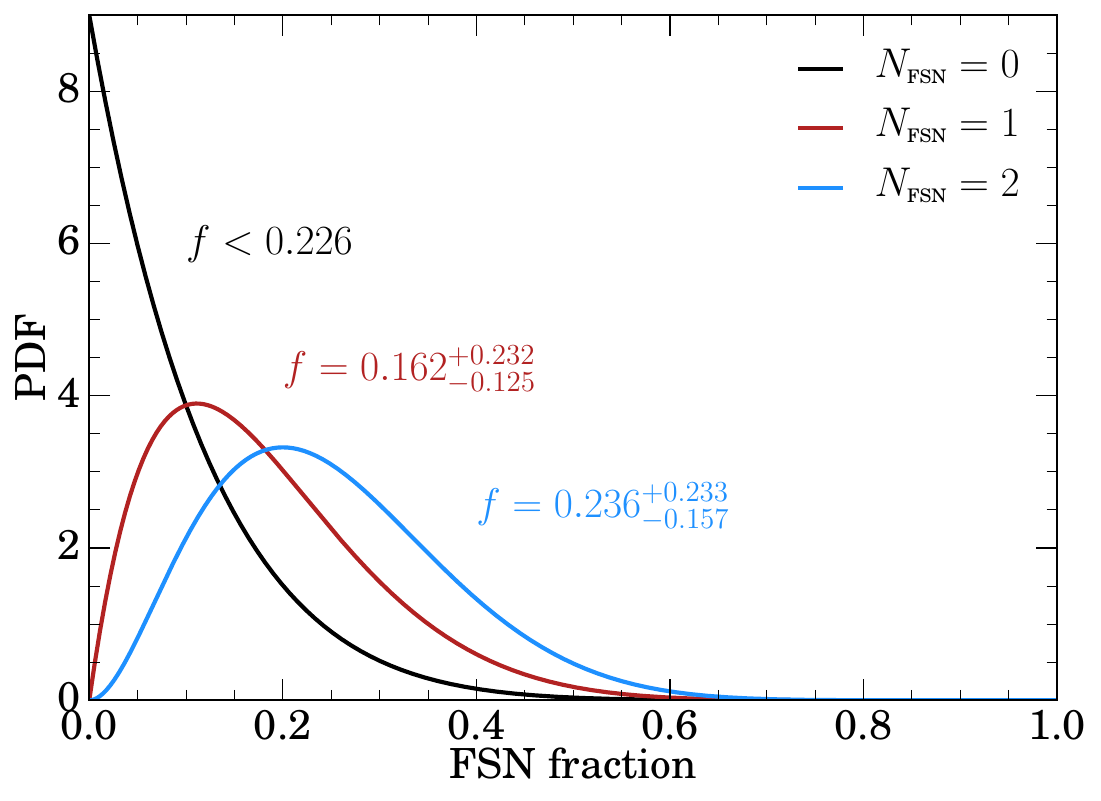}
\caption{PDF of failed SN fraction given zero (black), one (red), or two (blue) candidates.  For the one and two candidate PDFs, median fraction estimates are given, as well as the 90 per~cent confidence limits.  For the zero candidate PDF, a 90 per~cent upper limit is given.}
\label{fig:pdf}
\end{figure}

\begin{table}
\caption{Failed supernova/core-collapse fraction}
\label{tab:fraction}
\begin{center}
\begin{tabular}{cccc} \toprule
\multicolumn{1}{c}{$N_\text{FSN}$}	& Lower limit &  Median 	& Upper limit \\
\midrule
2 & 0.079 & 0.236 & 0.470 \\
1 & 0.037 & 0.162 &  0.394 \\ \midrule
0 & $-$ & $-$ & 0.226 \\
\bottomrule
\end{tabular}
\begin{flushleft}
\textit{Notes}: Limits are presented at the 90 per~cent confidence level.
\end{flushleft}
\end{center}
\end{table}

In Figure \ref{fig:pdf}, we present the PDFs for the failed SN fraction given $N_\text{SN} = 8$ and $N_\text{FSN} = 0,1,2$.  The $N_\text{FSN} = 2$ case assumes that both M101-OC1 and N6946-BH1 are true failed SNe.  The $N_\text{FSN} = 1$ case assumes that only N6946-BH1 is a failed SN.  For both cases, we calculate the median fraction as well as the 90 per~cent confidence bounds and present them in Table \ref{tab:fraction}.  For the $N_\text{FSN} = 1$ case, we calculate a median fraction\footnote{In A17b, the reported median fraction was 0.143, but this was actually the mode.  The median for $N_\text{SN} = 6$ and $N_\text{FSN} = 1$ is 0.201.} of 0.162 with a 90 per~cent confidence interval of $0.037 < f < 0.394$.  This is in line with estimates of $\sim$10--30 per~cent that are based on missing RSG progenitors and the Galactic BH mass function \citep{kochanek15}.  In comparison with the failed SN fraction reported in A17b using $N_\text{SN} = 6$, this measurement represents a shrinking of the confidence interval by 7 percentage points.  If all of the candidates are ultimately rejected, we generate an upper limit of $f < 0.226$ at 90 per~cent confidence, also presented in Table \ref{tab:fraction}. 

Future observations with the LBT are planned so as to continue the survey, detect new candidates, and constrain the failed SN fraction to better precision.  Parallel observations will also be obtained with \hst~and the soon-to-be-launched \textit{JWST} that will help analyze and confirm the candidates we have detected.  

\section{Acknowledgements}

Financial support for this work was provided by NSF through grant AST-1814440. This work is based on observations made with the Large Binocular Telescope. The LBT is an international collaboration among institutions in the United States, Italy and Germany. LBT Corporation partners are: The University of Arizona on behalf of the Arizona Board of Regents; Istituto Nazionale di Astrofisica, Italy; LBT Beteiligungsgesellschaft, Germany, representing the Max-Planck Society, The Leibniz Institute for Astrophysics Potsdam, and Heidelberg University; The Ohio State University, representing OSU, University of Notre Dame, University of Minnesota and University of Virginia.  This work also utilized observations made with the NASA/ESA \textit{Hubble Space Telescope}, obtained from the data archive at the Space Telescope Science Institute. STScI is operated by the Association of Universities for Research in Astronomy, Inc. under NASA contract NAS 5-26555. This work is based in part on observations made with the \textit{Spitzer Space Telescope}, which is operated by the Jet Propulsion Laboratory, California Institute of Technology under a contract with NASA.

\section*{Data availablity} 

The data underlying this article are available in the article and in its online supplementary material.


\bibliographystyle{mnras}
\bibliography{bibliography}


\section*{Appendix}

In Tables \ref{tab:n4736} and \ref{tab:m101}, we show abridged LBT light curve data of N4736-OC1 and M101-OC1 respectively.

\begin{table*}
\caption{N4736-OC1 light curve data}
\label{tab:n4736}
\begin{tabularx}{\textwidth}{ccccccccccccc} \toprule
MJD  & $m_R$ & $M_R$ & $m_R$ err & $\nu L_\nu(R)~[L\odot]$ & $m_V$ & $M_V$ & $m_V$ err & $\nu L_\nu(V)~[L\odot]$ & $m_B$ & ... & $m_U$ err & $\nu L_\nu(U)~[L\odot]$ \\ \midrule
54535.4 & 22.32 & -6.30 & 0.04 & 14318.39 & $-$ & $-$ & $-$ & $-$ & $-$ & ... & $-$ & $-$ \\
54859.4 & 22.38 & -6.24 & 0.05 & 13604.59 & 22.57 & -6.06 & 0.06 & 15894.51 & 23.19 & ... & 0.33 & 2512.2 \\
54916.4 & 22.22 & -6.40 & 0.06 & 15680.30 & $-$ & $-$ & $-$ & $-$ & 22.95 & ... & 0.26 & 6450.57 \\ 
... & ... & ... & ... & ... & ... & ... & ... & ... & ... & ... & ... & ... \\
58258.3 & 23.53 & -5.09 & 0.07 & 4715.36 & 24.19 & -4.44 & 0.22 & 3577.2 & 25.07 & ... & 0.32 & 1781.09 \\
\bottomrule
\end{tabularx}
\begin{flushleft}
\textit{Notes}:  Full data is available as supplementary material as a machine-readable table.
\end{flushleft}
\end{table*}

\begin{table*}
\caption{M101-OC1 light curve data}
\label{tab:m101}
\begin{tabularx}{\textwidth}{ccccccccccccc} \toprule
MJD  & $m_R$ & $M_R$ & $m_R$ err & $\nu L_\nu(R)~[L\odot]$ & $m_V$ & $M_V$ & $m_V$ err & $\nu L_\nu(V)~[L\odot]$ & $m_B$ & ... & $m_U$ err & $\nu L_\nu(U)~[L\odot]$ \\ \midrule
54534.5 & 22.40 & -6.66 & 0.04 & 19975.47 & 22.54 & -6.52 & 0.04 & 24404.58 & 22.73 & ... & $-$ & $-$ \\
54535.5 & 22.42 & -6.64 & 0.03 & 19613.66 & 22.52 & -6.54 & 0.03 & 24776.21 & 22.72 & ... & 0.03 & 38315.53 \\
54536.5 & 22.40 & -6.66 & 0.02 & 20022.80 & 22.54 & -6.52 & 0.04 & 24371.69 & 22.73 & ... & $-$  & $-$ \\ 
... & ... & ... & ... & ... & ... & ... & ... & ... & ... & ... & ... & ... \\
58838.5 & 23.14 & -5.92 & 0.05 & 10092.65 & 23.37 & -5.70 & 0.06 & 11376.49 & 23.95 & ... & 0.20 & 4495.01 \\
\bottomrule
\end{tabularx}
\begin{flushleft}
\textit{Notes}: Full data is available as supplementary material as a machine-readable table.
\end{flushleft}
\end{table*}

\label{lastpage}
\bsp	
\end{document}